\documentclass[twocolumn,showpacs,preprintnumbers]{revtex4}
\usepackage[]{graphicx}
\begin{document}

\title{Quasi symplectic integrators for stochastic differential equations}
\author{R. Mannella}

\affiliation{Dipartimento di Fisica and INFM, UdR Pisa, Universit\`a degli
Studi di Pisa, Via Buonarroti 2, 56100 Pisa, Italy}

\date{\today} 

\begin{abstract}
Two specialized algorithms for the numerical integration of the 
equations of motion
of a Brownian walker obeying detailed balance are introduced. 
The algorithms
become symplectic in the appropriate limits, and reproduce
the equilibrium distributions to some higher order in the integration
time step. Comparisons with other existing integration schemes are carried out
both for static and dynamical quantities.
\end{abstract}
\maketitle

\section{Introduction}
Stochastic processes are well known to be at the heart of many 
physical systems~\cite{cup}. Several approaches have hence been developed
to understand the dynamics which is realized given specific models: among
others, one of the most used ones is Monte-Carlo integration of the equations 
of motion.

The literature on the numerical integration of stochastic differential
equations is huge: we will limit here to mention a couple of classical 
citations widely used in the physics 
community~\cite{kloeden,sancho99,milnstein97}. Additional comments and
references can be found in~\cite{mann02}. The numerical algorithms
presented in these works are general and can be applied to basically
any flow; however, they might not be the optimal ones for cases when 
additional
information about the details of the system under study are available.

An important class for which dedicated algorithms can be derived
is given by the equations of motion
\begin{eqnarray} 
\dot x &=& v \nonumber \\
\dot v &=& -\gamma v + F(x) + \xi(t) \label{eq1} 
\end{eqnarray}
where $\xi(t)$ is a random Gaussian noise, with zero average and
standard deviation 
$$ \langle \xi(t) \xi(s) \rangle = 2 \gamma T \delta (t-s).$$
In the following, we will also use $V(x)$, defined as 
$F(x) \equiv - V'(x)$. Note that although we are dealing here with only
one Brownian walker, the algorithm we are going to show can be easily extended
to the case when $x$ and $v$ are vectors, and $F_i(x)$, the force
acting on the $i$-th walker, is a function of all other walkers.

The above equation is commonly found in the liquid state literature
(for numerical schemes appropriate in the integration of
the Brownian dynamics of a liquid, see among others
\cite{allen,branka,forbert,qiang,paul})
and some algorithms have been proposed, over the years, for its
numerical integration.

To date, perhaps the most widely used algorithms for the 
integration of Eq. \ref{eq1} 
are the ones derived in 
\cite{allen}, where two algorithms have been proposed (see also references
therein): we will benchmark against one of them, and to this end,
we will briefly review them below. Note that
the system of Eq. \ref{eq1} becomes symplectic when 
$\gamma \rightarrow 0$ and, until some recent 
works~\cite{miln02a,miln03a,miln03b}, this symplectic nature was not really
exploited in deriving numerical schemes.

The approach we will follow is to derive numerical algorithms having
in mind two requirements: (i) the algorithm should become
symplectic in the deterministic ($T=0$) and frictionless ($\gamma=0$) 
limit; and (ii) the numerical algorithm should reproduce as closely
as possible the equilibrium distribution, when it is defined, of 
the system given
by Eq. \ref{eq1}. As we will see below, to the best of our 
knowledge either the former or the latter requirements have been enforced
in the derivation of numerical schemes, but never both of them.
The algorithms introduced here will improve both the 
algorithms of~\cite{allen} and of~\cite{miln03b}.

\section{Brief review of the benchmark algorithms and some definitions}

To assess how well each algorithm is performing, we 
start from the knowledge that
for $V(x)$ which are bounded from below, 
Eq. \ref{eq1} leads to an equilibrium distribution $P(x,v)$ for the
variables $x$ and $v$ of the form
\begin{equation}
P(x,v) = N \exp{\left\{-\left(v^2/2 + V(x)\right)/T\right\}}.
\label{eq2}
\end{equation}
where $N$ is a normalization constant. We are going to 
compare the exact theoretical 
equilibrium distribution
to the equilibrium distribution obtained from the simulations.
It is possible, in principle,
to check theoretically which is the equilibrium distribution which
is expected integrating using a given numerical
scheme, following~\cite{batrouni}: suppose we have a 
numerical scheme of the form
$$x_i(t+h) = x_i(t) + h W_i (x_i,\xi) $$
then the probability distribution of $x_i$ satisfies
\begin{equation}
P(x_i,t+h)-P(x_i,t)=\sum_{n=1}^{\infty}\sum_{x_{i}}
\frac{\partial}{\partial x_{i}} \ldots \frac{\partial}{\partial x_{n}}
K_{1 \dots n}P(x_i,t)
\label {batr}
\end{equation}
with
$$
K_{1 \dots n} \equiv (-1)^{n}\frac{1}{n!} \langle W_{1} \ldots
W_{n}\rangle_{\xi},
$$
where $\langle \ldots \rangle_\xi$ means averaging over the noise
realizations.
In general, for systems in detailed balance at temperature $T$, 
$$
P(x_i,t=\infty)_{sim} = P(x_i,\infty)_{true} \times
\exp{\left(\sum_{n=1}^{\infty}
h^{n}S_{n}/T\right)}
$$
where $P_{sim}$ is the equilibrium distribution generated in the
simulations, and $P_{true}$ is the theoretical equilibrium distribution. 
Given the explicit form of the numerical scheme, the various 
$K_{1 \dots n}$ can be computed: applying Eq. \ref{batr}, and expanding
the r.h.s. of Eq. \ref{batr}
in the small parameter $h$, assuming (equilibrium) that
$P(x_i,t+h)-P(x_i,t)=0$, 
we can derive the equations satisfied by
the $S_i$.  The first nonzero 
$S_i$ yields the correction to the true equilibrium distribution generated
by the numerical scheme.

Let us show how to use Eq. \ref{batr} taking one of
the algorithms of~\cite{allen}. This algorithm 
integrates Eq. \ref{eq1} using the
prescription
\begin{eqnarray}
 & x(t+h) = x(t) + c_1 h v (t) + c_2 h^2 F(x(t)) + \eta_1 &\nonumber \\
 & v(t+h) = c_0 v(t) + c_1 h F(x(t)) + \eta_2 &\label{liq1}
\end{eqnarray}
where
$$ c_0 = e^{-\gamma h} \;\;\;
c_1 = \frac{1 - c_0}{\gamma h} \;\;\;
c_2 = \frac{1-c_1}{\gamma h} $$
and where $\eta_1$ and $\eta_2$ are two random Gaussian variables with
zero average and moments 
$$\langle \eta_1^2\rangle = \frac{Th}{\gamma} \left( 2 - \frac{3 - 4
e^{-\gamma h} + e^{-2 \gamma h}}{\gamma h}\right) $$
$$\langle \eta_2^2 \rangle = T \left(1-e^{-2\gamma h}\right) $$
$$\langle \eta_1 \eta_2 \rangle  = \frac{T}{\gamma}\left(1-e^{-\gamma
h}\right)^2$$
The algebra to derive the correction to the equilibrium distribution 
induced by the
numerical scheme in the general case and for a given high order integration 
scheme can be formidable; however, for a flow like Eq. \ref{eq1} 
 and for the scheme given
by Eq. \ref{liq1}, the algebra is manageable. Assuming that 
(see Eq. \ref{eq2}, with $S \equiv S_1$)
$$ P(x,v) =N \exp\left\{-\left[v^2/2 + V(x)+h S(x,v)\right]/T\right\}, $$
plugging the scheme of Eq. \ref{liq1} into Eq. \ref{batr}, 
we have that $S(x,v)$ satisfies the partial differential equation
$$\frac{\partial^2 S(x,v)}{\partial v^2} - \frac{v}{T\gamma}
\frac{\partial S(x,v)}{\partial x} - \left(\frac{F(x)}{T\gamma} -
\frac{v}{T}\right) \frac{\partial S(x,v)}{\partial v} $$
$$ - \frac{v^2}{2\gamma T} F'(x) - \frac{1}{2  \gamma} F'(x) = 0 $$
This implies that this algorithm fails to reproduce the correct
equilibrium distribution at $O(h)$ in the exponent. It is possible,
for the case when $V(x)= \omega^2 x^2/2$, to solve this partial 
differential equation, obtaining the numerical equilibrium
distribution at lowest order in $h$, which reads
$$P(x,v) = N \exp \left\{-\left[v^2/2 + \omega^2 x^2/2\right]/\hat{T}
\right\} $$
with
$$ \hat{T} = \frac{T}{1+ \frac{\omega^2 h}{2 \gamma}} $$
i.e. the numerical equilibrium distribution is similar to the correct one, but
with a renormalization of the temperature. In particular, this
effective temperature (the temperature ``simulated'' by the algorithm)
goes to zero in the limit $\gamma \rightarrow 0$. 
In \cite{allen} it is acknowledged 
that the algorithm does not
work well in this limit, although no formal proof is provided.

To overcome the problem with the case of small $\gamma$, in
\cite{allen} a second algorithm is proposed, which reads
\begin{eqnarray}
x(t+h) &=& x(t) + c_1 h v(t) + c_2 h^2 F(x(t)) + \eta_1 \nonumber \\
v(t+h) &=& c_0 v(t) + (c_1-c_2)h F(x(t)) + \nonumber \\
       & & c_2 h F(x(t+h)) + \eta_2.
\label{liq2}
\end{eqnarray}

Using Eq. \ref{batr} to evaluate the correction to the true equilibrium
distribution generated by this algorithm, we find that the 
contribution $S_1$ vanishes, and we are left with the term $S_2$.
In other words, this algorithm reproduces the correct equilibrium 
distribution at $O(h)$, but there are still corrections $O(h^2)$ 
in the exponent. The algorithm given in Eq. \ref{liq2} is the
reference algorithm which we propose to improve in the next section.
We will refer to this algorithm as ``Li'' (from \textit{Li}quid) 
in the following.

\section{Quasi symplectic algorithms}

A symplectic algorithm is a numerical scheme which attempts to preserve
the 2-forms $dq_i \times dp_i$ during the integration of a Hamiltonian 
flow. The quantity $q_i$ 
is a generalized coordinate, and $p_i$ is the corresponding conjugate
momentum. A nice introduction to the symplectic integration can be 
found in~\cite{yoshida,serna}. 
Given the Hamiltonian $H(q_i,p_i)$, and the equations of motion
$$ \dot q_i = \{q_i,H\} \;\;\;\;\; \dot p_i = \{p_i,H\}, $$
a symplectic integrator will in practice 
conserve some quantity $\hat H$, which in 
general reads
$$\hat H = H + h^n G(p_i,q_i) $$
where $h$ is the integration time step, and $G$ is a function which 
depends on the numerical scheme used for the integration.
The problem of Hamiltonian flows in the presence of fluctuations has
been addressed also in~\cite{miln02a,miln03a}, whereas quasi
symplectic schemes were derived in~\cite{miln03b} (see also below, when
various comparisons are carried out). A preliminary account of
the material of this section can be found in~\cite{rm.rm1}.

Given that we are interested here in the integration of Eq. \ref{eq1},
we start from the symplectic integration of Hamiltonians which
are separable and quadratic in the velocities. 
There are very many different possible
symplectic schemes: however, having in mind that we are seeking a 
scheme which should be used in the integration of a stochastic differential
equation, we restrict ourselves to considering a scheme in the form
$$ q(i) = q(i-1) + h  a_i p(i-1) $$
$$ p(i) = p(i-1) + h  b_i F(q(i)) $$
for $i$ between 1 and $N$, where $q(0)=q(t=0)$, 
$q(N)=q(t=h)$, etc., and $h$ is the integration time 
step in the simulations. The coefficients $a(i)$ and $b(i)$ are chosen
as to minimize, in some sense, the quantity $G(p,q)$.

The lowest possible symplectic algorithm one can write to integrate
Eq. \ref{eq1} following this
approach reads when $\gamma=T=0$ (this scheme is also known as ``leap frog'')
\begin{eqnarray}
\tilde x & = & x(t) + \frac{h}{2} v(t) \nonumber \\
v(t+h)   & = & v(t) + h F(\tilde x) \nonumber \\
x(t+h)   & = & \tilde x + \frac{h}{2} v(t+h) 
\end{eqnarray}
where $x$ is the position and $v$ is the velocity. This scheme conserves
the quantity $H-h^3 (v F F' + v^3 F''/6)/4$, where $H \equiv v^2/2+V(x)$.
It is then possible to reintroduce both the dissipation ($-\gamma v$)
and the noise, writing the tentative scheme
\begin{eqnarray}
\tilde x & = & x(t) + \frac{h}{2} v(t) \nonumber \\
v(t+h)   & = & c_2\left[ c_1 v(t) + h F(\tilde x) + d_1 \eta 
\right]\nonumber \\
x(t+h)   & = & \tilde x + \frac{h}{2} v(t+h)
\label{sym1} 
\end{eqnarray}
where $\eta$ is a Gaussian variable, with standard deviation one and
average zero.  We use again Eq. \ref{batr}, and, imposing that the 
term $O(h)$ in the exponent (i.e. the term $h S_1$) vanishes, we find
that the unknown arbitrary quantities $c_1$, $c_2$ and $d_1$ read
\begin{eqnarray}
c_1  & = & 1 - \frac{\gamma h}{2} \nonumber \\
c_2  & = & \frac{1}{1+\gamma h/2} \nonumber \\
d_1  & = & \sqrt{2 T \gamma h}.
\end{eqnarray}
Although we will carry out more extensive comparisons further down, let us
briefly compare this scheme to the scheme of Eq. \ref{liq2}. The present
scheme is by construction well behaved in the limit of 
$\gamma \rightarrow 0$; it has
the same accuracy in computing the equilibrium distribution as of 
Eq. \ref{liq2}; 
but it requires only one random deviate per integration
time step (as opposed to two deviates for the scheme of Eq. \ref{liq2}), so it 
will run faster. In the following, we will refer to the algorithm
of Eq. \ref{sym1} as ``SLO'' (Symplectic Low Order).

Looking at the structure of the previous algorithm, we can try to
derive an algorithm of higher order. In the derivation of Eq. \ref{sym1},
when we applied Eq. \ref{batr}, given the number of unknown quantities
we could only impose that the term $O(h)$ in the exponent disappeared.
If we could somehow increase the number of unknown quantities when
applying Eq. \ref{batr}, while
keeping the algorithm simple, we might be able to make the terms
$O(h^2)$ in the exponent disappear.
We start combining two
steps, each one of the form of Eq. \ref{sym1}, done with 
an integration time step $h/2$,
\begin{eqnarray}
x_1 & = & x(t) + \frac{h}{4} v(t) \nonumber \\
v_1   & = & c_2\left[ c_1 v(t) + h F(x_1) +\sqrt{\gamma T h} 
(a_1\eta_1 + a_2 \eta_2)
\right]\nonumber \\
x_2 & = & x_1 + \frac{h}{2} v_1 \nonumber \\
v(t+h)   & = & c_2\left[ c_1 v_1 + h F(x_2) +\sqrt{\gamma T h} 
(b_1 \eta_1 + b_2 \eta_2)
\right]\nonumber \\
x(t+h)   & = & x_2 + \frac{h}{4} v(t+h) 
\label{sym2}
\end{eqnarray}
where $c_1 = (1-\gamma h/4)$ and $c_2=1/(1+\gamma h/4)$. Here, $\eta_1$ and
$\eta_2$ are two random Gaussian variables of standard deviation one and 
average zero. The idea is now to choose the coefficients $a$'s and $b$'s 
in such a way as to annihilate $S_1$, and possibly minimize $S_2$. This is
done using Eq. \ref{sym2} in Eq. \ref{batr}, which results in a number of
algebraic equations for $a_i$ and $b_i$. The algebra, although
straightforward, is cumbersome and we will simply report here the
results.
For a given $a_2$, the following choice for the other three parameters
will ensure that $S_1$ vanishes identically:
$$b_2 = -\frac{a_2}{7}+\frac{2\sqrt{14- 12 a_2^2}}{7} $$
$$b_1 = -\frac{\sqrt{7+282 a_2^2 + 24 a_2 \sqrt{14-16 a_2^2}}}{\sqrt{42}}$$
$$a_1 = \frac{\sqrt{42} b_1 (-7\sqrt{2}+6\sqrt{2}a_2^2+24 a_2 \sqrt{7-6 a_2^2})}{\sqrt{3} (-14+588 a_2^2)}.$$

As function of $a_2$, we can now write the equations satisfied by $S_2$:
we find that $S_2$ vanishes for a particular choice of the parameter $a_2$. 
Summarizing the numerics, the set of $a$'s and $b$'s which simultaneously
makes $S_1$ and $S_2$ vanish are:
\begin{eqnarray}
   a_1 &=& -1.0691860043307065 \ldots  \nonumber \\
   a_2 &=& -0.1533230407019893 \ldots  \nonumber \\
   b_1 &=&  0.3044913128854065 \ldots  \nonumber \\
   b_2 &=& -1.0363164126095790 \ldots . 
\label{sym3}
\end{eqnarray}
The conclusion is that the algorithm given by Eqs. \ref{sym2} and
\ref{sym3} is symplectic in the limit $\gamma \rightarrow 0$
(conserving the quantity $H-h^3 (v F F' + v^3 F''/6)/16$), whereas for
a finite $\gamma$ it reproduces the correct equilibrium
distribution with an error $O(h^3)$ in the exponent. We will refer to this
algorithm as to ``SHO'' (Symplectic High Order).

In the following, we will use also the Heun algorithm to carry out the
various comparisons. To make this paper as self contained as possible,
we recall here that the Heun algorithm for a system like the one in
Eq. \ref{eq1} is given by the prescription:
\begin{eqnarray}
 x_1 &=& x(t) + h v(t) \nonumber \\
 v_1 &=& v(t) - h \gamma v(t) + h F(x(t)) + \sqrt{2 \gamma T h} \eta\nonumber \\ 
 x_2 &=& x(t) + h v_1 \nonumber \\
 v_2 &=& v(t) - h \gamma v_1 + h F(x_1) + \sqrt{2 \gamma T h} \eta\nonumber \\
 x(t+h) &=& \frac{x_1+x_2}{2} \;\;\;\;  v(t+h) = \frac{v_1+v_2}{2} \nonumber
\end{eqnarray}
The Heun algorithm does not make use of the (quasi) symplectic nature of
the flow: we expect that it will not fare too well in the limit of small
$\gamma$. We recall that it is known \cite{mann02} 
that the equilibrium distribution generated by the
Heun algorithm is accurate up to $o(h^2)$ in the exponent. We will refer to
the Heun scheme as to ``He''.

We will also compare our algorithms to the quasi symplectic
algorithms of~\cite{miln03b}: we should mention here that really the latter
are weak integration schemes (for a definition of weak and strong integration
schemes, see~\cite{kloeden}), hence they are bound to give worse results than
the other schemes when, as we do, average quantities are considered. 
The two algorithms considered integrate with the
prescriptions (\cite{miln03b} should be consulted for more details):
\begin{itemize}
\item[MT1:] \begin{eqnarray}
  & x(t+h) & = x(t) + h v(t+h) \nonumber \\
  & v(t+h) & = v(t) - h V'(x(t+h))-h \gamma v(t+h) \nonumber \\
  &        & + \sqrt{2 T h \gamma} \eta
  \label{mt1}
  \end{eqnarray}
where $v(t+h)$ and $x(t+h)$ should be found recursively.
\item[MT2:] \begin{eqnarray}
  & v(t+h) & = (1-\gamma h) ( v(t) - h V'(x(t)) + 
\sqrt{2 T h \gamma} \eta) \nonumber \\
  & x(t+h) & = x(t) + h (v(t) - h V'(x(t))) 
  \label{mt2}
\end{eqnarray}
\end{itemize}
The random variables $\eta$ take the values $\pm 1$. 
These variables are faster to generate
than a Gaussian variable, hence these algorithms will run faster, 
allowing for a smaller
integration timestep to compensate for less accuracy when averaged 
quantities are considered. However, having said this, if we used 
Eq. \ref{batr} to asses these two algorithms, we would find
that they both have a correction to the equilibrium
distribution $O(h)$ in the exponent: these will reflect in the 
numerical experiments, as we will comment below.

Finally, it should be noted that given the structure of the Hamiltonian 
in the limit $T \rightarrow 0$ and $\gamma \rightarrow 0$, 
which is $H=v^2/2+V(x)$, the equilibrium distribution for
Eq. \ref{eq1} can also be written as
$$P(x,v) \propto \exp{-H/T}.$$
At first sight, it may appear that the request of a symplectic
integration scheme is redundant, once we made sure that the 
``correct'' equilibrium distribution is generated in the numerical 
integration. This is not right: the limit $T \rightarrow 0$
is singular, hence a symplectic form for the numerical scheme 
can (and should) be imposed as an
additional condition.

\begin{figure}[htb]
\includegraphics[width=3 in]{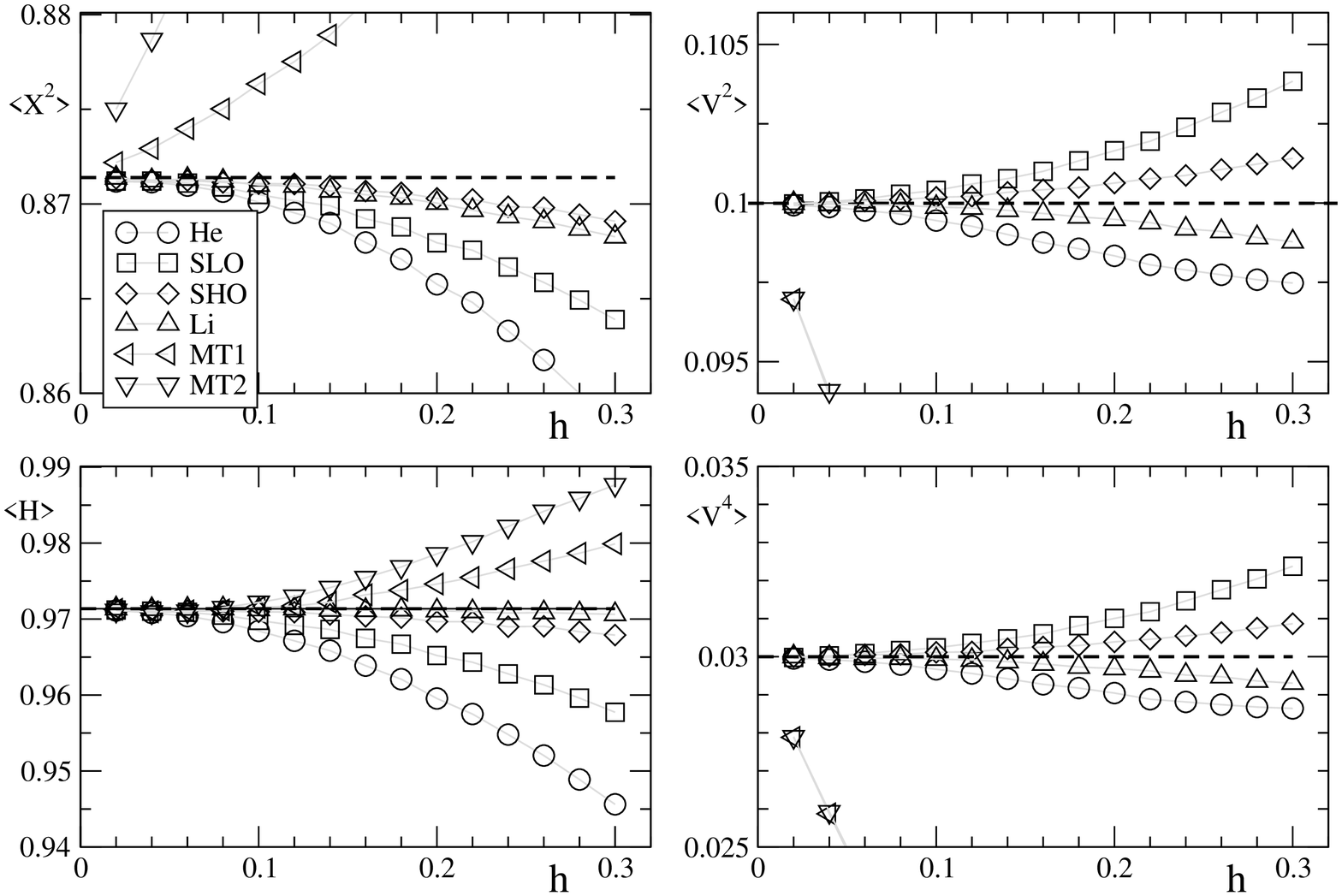}
\caption{Result of simulations for different integration schemes, as
function of the integration time step $h$, for the system in
Eq. \protect{\ref{sys1}}. Various moments are considered (see
text for details), taking $T=0.1$ and $\gamma=1$}
\label{fig1}
\end{figure}

\begin{figure}[htb]
\includegraphics[width=3 in]{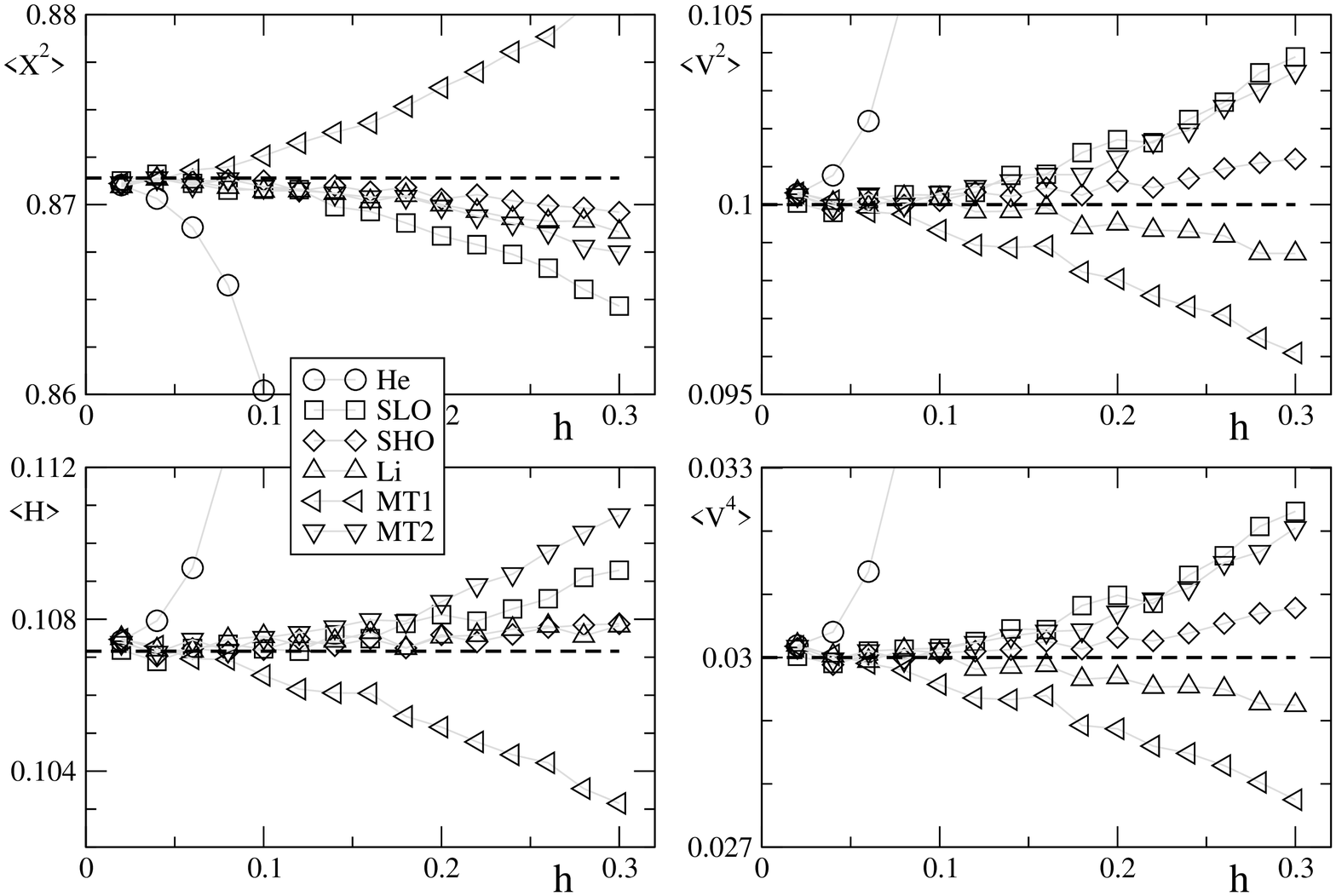}
\caption{Result of simulations for different integration schemes, as
function of the integration time step $h$, for the system in
Eq. \protect{\ref{sys1}}. Various moments are considered (see
text for details), taking $T=0.1$ and $\gamma=10^{-2}$}
\label{fig2}
\end{figure}

\section{Numerical experiments}

We compare now the results of applying the algorithms described previously
to the integration of two prototype stochastic differential equations. Let us
first consider how the different algorithms reproduce the equilibrium
properties: to this end, we integrate the equations of motion, 
and compute some equilibrium momenta, which are then
compared to the theoretical ones.

The first system studied is given by
\begin{eqnarray}
\dot x &=& v \nonumber \\
\dot v &=& -\gamma v - V'(x) + \sqrt{2 \gamma T} \eta \\
V(x) &=& x^4/4 - x^2/2 \nonumber
\label{sys1}
\end{eqnarray}

For this system we fixed the noise intensity to $T=0.1$, and carried out
the numerical integration for two different values of the damping coefficient 
$\gamma$ and for the different integration schemes. 
The results are summarized in Fig.~\ref{fig1} (for $\gamma=1$) and in 
Fig.\ref{fig2} (for $\gamma=10^{-2}$). The
quantity $\langle H \rangle$ is defined as
$\langle H \rangle \equiv \langle v^2/2+V(x) \rangle$. In all figures, the
results of the digital simulations are shown by symbols with a gray
straight line as guide to the eye; the bold dashed line is the
expected (theoretical) value for the quantity considered. For the number
of averages considered, the statistical error is much smaller than 
(order of) the
symbols for the larger (smaller) damping.

Let us comment the
results. It is evident that the Heun method (He in figures) is not very 
appropriate for the smaller damping considered (Fig.~\ref{fig2}). Even for
the larger damping (Fig.~\ref{fig1}), the Heun algorithm is typically
outperformed by the Symplectic Low Order scheme (SLO, Eq. \ref{sym1}); 
note that the SLO
is fairly faster than He, given that it requires only one evaluation of
the deterministic force for each integration time step.

The algorithms MT1 and MT2, as expected, do not work well for the larger 
damping considered, and become more accurate as the damping is reduced: it should be
noted that for this case, MT2 seems to be more accurate than MT1 for a
given integration time step: considering that MT2 is much faster than MT1,
the conclusion seems to be that MT2 ought to be preferred, between these
two schemes. Note also that the error on the moments for these two
schemes seems to grow linearly with the integration time step $h$, which
is related to the $O(h)$ error in the exponent which was mentioned 
in the previous section.

The Li algorithm is not particularly accurate when the $x^2$ moments are 
considered: for both values of the damping, SLO gives more accurate
results for this moment, in the whole $h$ region. The Symplectic High
Order (SHO) is the algorithm which gives the most accurate results for the
$x^2$ moment, and results comparable or better than to the one obtained 
with Li for the
$v^2$ and $v^4$ moments. It is only when
$\langle H \rangle$ is considered, and for the larger damping, 
 that Li seems to be more accurate than
SHO. However, care is necessary in drawing conclusions from 
$\langle H \rangle$: looking for instance at Fig.~\ref{fig1}, we note
that Li underestimates both $v^2$ and $x^2$: recalling the structure of the
potential $V(x) = -x^2/2+x^4/$, it is clear that these two underestimates 
tend to cancel out, leading to a $\langle H \rangle$ closer to the theoretical
one, but only by virtue of a coincidental cancellation.

\begin{figure} [htb]
\includegraphics[width=3 in]{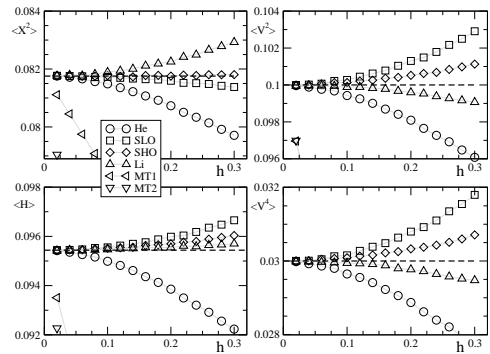}
\caption{Result of simulations for different integration schemes, as
function of the integration time step $h$, for the system in
Eq. \protect{\ref{sys2}}. Various moments are considered (see
text for details), taking $T=0.1$ and $\gamma=1$}
\label{fig3}
\end{figure}

\begin{figure} [htb]
\includegraphics[width=3 in]{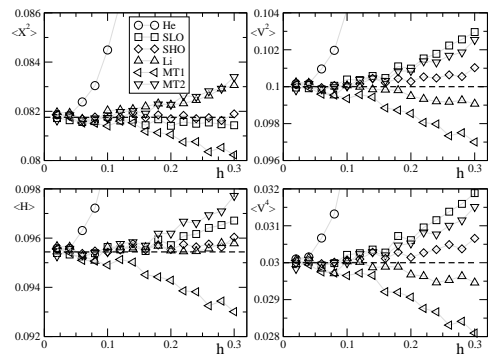}
\caption{Result of simulations for different integration schemes, as
function of the integration time step $h$, for the system in
Eq. \protect{\ref{sys2}}. Various moments are considered (see
text for details), taking $T=0.1$ and $\gamma=10^{-2}$}
\label{fig4}
\end{figure}
The second system studied is similar to the first one:
\begin{eqnarray}
\dot x &=& v \nonumber \\
\dot v &=& -\gamma v - V'(x) + \sqrt{2 \gamma T} \eta \\
V(x) &=& x^4/4 + x^2/2 \nonumber
\label{sys2}
\end{eqnarray}
the only difference with the system of Eq. \ref{sys1} 
being that now the potential is monostable.

The result of the computer experiments are summarized in Figs. \ref{fig3}
and \ref{fig4}. The comments parallel the comments we already made 
for the system of Eq. \ref{sys1}. Heun (He) is the least accurate 
scheme for small damping, although the error on the moments is 
quadratic on $h$ (a signature of an $O(h^2)$ error in the 
equilibrium distribution). MT1 and MT2 perform better at 
smaller damping, with an error on the moments which is roughly
linear in the integration time step. SLO does better than
both He and MT1, MT2, and for both damping considered, being as
fast (if not faster) than both schemes. When the $x^2$ is considered, Li 
appears to perform worse than SLO. SHO outperforms Li: only
when $H$ is considered, Li seems to be more accurate than SHO, but
again only by virtue of a cancellation (again, between
$\langle x^2\rangle$ and $\langle v^2\rangle$).

\begin{figure} [htb]
\includegraphics[width=3 in]{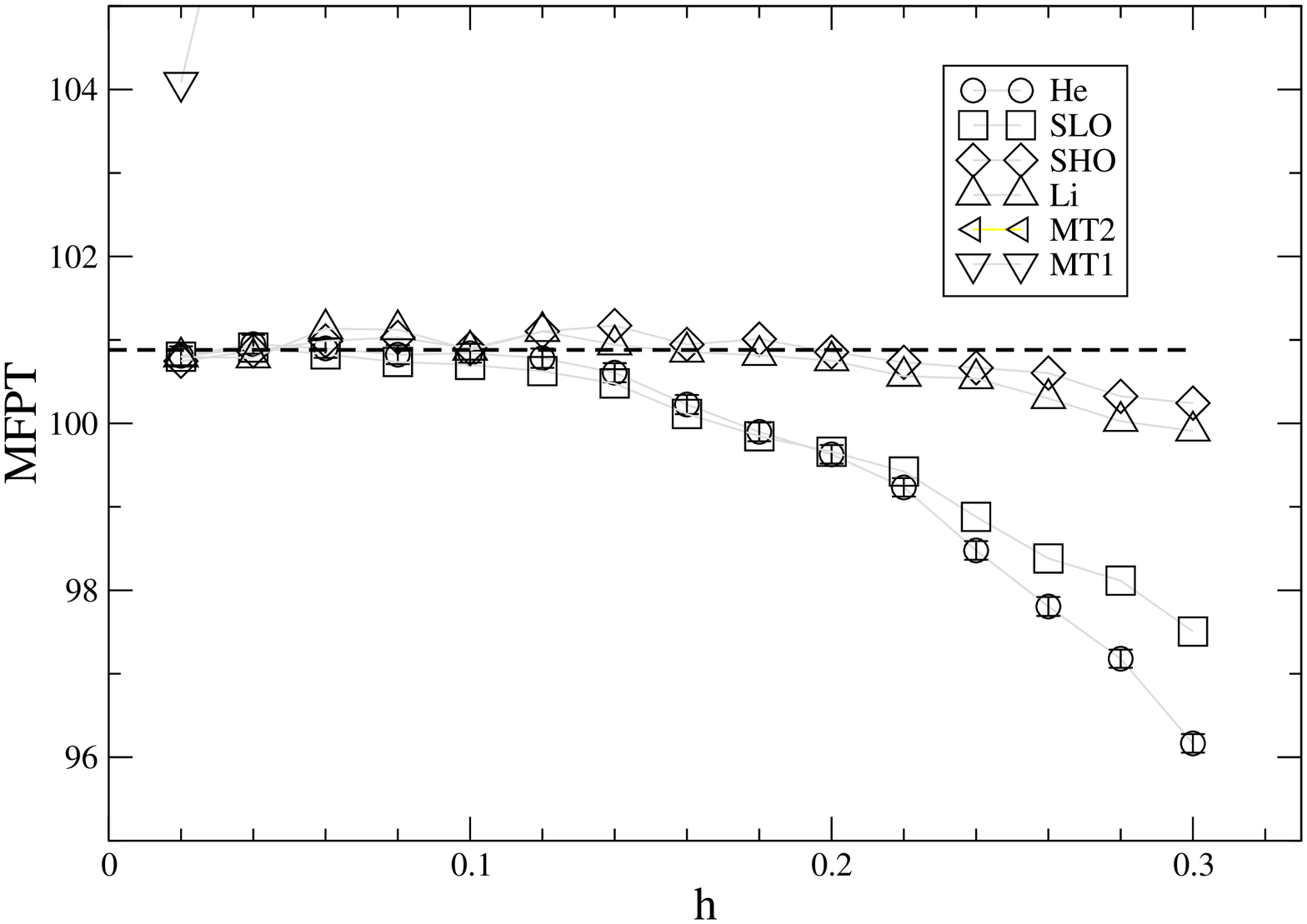}
\caption{Result of simulations for different integration schemes, as
function of the integration time step $h$, for the system in
Eq. \protect{\ref{sys1}}. The Mean First Passage Time between the
minima is considered (see
text for details), taking $T=0.1$ and $\gamma=1$}
\label{fig5}
\end{figure}

\begin{figure} [htb]
\includegraphics[width=3 in]{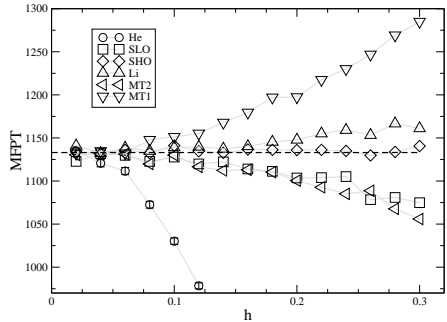}
\caption{Result of simulations for different integration schemes, as
function of the integration time step $h$, for the system in
Eq. \protect{\ref{sys1}}. The Mean First Passage Time between the
minima is considered (see
text for details), taking $T=0.1$ and $\gamma=10^{-2}$}
\label{fig6}
\end{figure}

We turn now to some numerical experiments to assess how the various 
algorithms perform when dynamical quantities are considered. Using
the system of Eq. \ref{sys1}, we computed the Mean First Passage
Times (MFPT) to go from one of the minima to the other one (the minima
are located at $x=\pm 1$): the results of the simulations are
summarized in Figs. \ref{fig5} and \ref{fig6}. Given the fairly 
large value of the noise intensity ($T=0.1$), there is no theory
available to compute an ``exact'' MFPT for comparison with the simulated MFPTs.
To have some reference, we took as reference value the average of
the MFPT obtained via
the algorithms Li and SHO for the three smallest $h$ values used in the
simulations, and drew the dashed line at this value. The 
expected statistical error, due to the finiteness of the sample used,
is of the order of the symbol size.
For the larger damping considered (Fig. \ref{fig5}, $\gamma=1$),
He and SLO perform in a similar way. MT1 and MT2 give unreasonable values
for the MFPT (only one point for MT1 is actually on the graph, for the
smallest $h$ considered: all other points for both algorithms are
outside the MFPT range considered). Li and SHO perform in a similar
way, giving results closer to the ``correct'' MFPT throughout the $h$ range 
considered, with a slightly
better agreement shown by SHO for larger $h$'s.
The situation is more interesting when a smaller damping is considered 
(Fig. \ref{fig6}, $\gamma=10^{-2}$). While showing an error which 
grows only quadratically in $h$, clearly He is the algorithm which 
performs worse. MT1 and MT2 now give more reasonable results, and they
yield MFPT comparable to the ones obtained using SLO. SHO outperforms
Li, giving MFPT closer to the ``exact'' ones 
over the whole $h$ range considered.
Li on the other hand seems to give results which are roughly equivalent
to the ones obtained using MT1, MT2 or SLO.

We would like to conclude noting that the numerical experiments were done
stretching the algorithms into parameter regions which are somehow 
extreme: the typical time scale for the potential 
considered is around 0.5 (the oscillation frequency around the minima
for Eq. \ref{sys1}) or around 1 (the larger $\gamma$ considered, and
the oscillation frequency for the potential of Eq. \ref{sys2}), and yet
an algorithm like SHO is able to integrate up to integration 
time steps $h$ order
of 0.3, with corrections to the moments or to the MFPTs 
which are smaller than, or at worse
order of 1\%: in our opinion, these are remarkable results,
particularly when the flatness of the MFPT computed with SHO
in Fig. \ref{fig6} is considered.

\section{Conclusions}

We introduced two algorithms for the numerical integration of the 
equations of motion of a Brownian walker. The features of these
algorithms are that they become symplectic when the damping
on the Brownian walker is taken to be zero, and give the correct
equilibrium distribution to some higher order in the integration
time steps for a finite damping and temperature. 
This, in turn, leads to more accuracy when dynamical
quantities are considered (like the MFPT). Possible applications of
these algorithms, beside the mentioned generic integration of the dynamics
in the liquid state \cite{allen}, are in the integration of the
dynamical equations of ions moving in and around ionic 
channels \cite{moy}: here the speed up provided by algorithms which
are stable for fairly large time steps may help in improving
current simulations.

\end{document}